\documentclass[sort&compress, twocolumn, 5p]{elsarticle}

\usepackage{amsmath}
\usepackage{graphicx}
\usepackage{wasysym}

\def\HOmI{\mbox{HOMO-1}}
\def\HOmII{\mbox{HOMO-2}}
\def\intensity[#1][#2]{\mbox{$#1\times10^{#2}$~W/cm$^2$}}

\newcommand{\rl}{\rule[-0.25mm]{0mm}{5mm}}

\begin{document}

\title{Water Molecules in Ultrashort Intense Laser Fields}

\author{Simon Petretti}
\author{Alejandro Saenz\corref{cor1}}
\ead{saenz@physik.hu-berlin.de}
\cortext[cor1]{Corresponding author}
\address{
    AG Moderne Optik,
    Institut f\"ur Physik,
    Humboldt-Universit\"at zu Berlin,
    Newtonstr. 15, D\,-\,12\,489 Berlin, Germany
}

\author{Alberto Castro}
\address{
     Institute for Biocomputation and Physics of Complex Systems (BIFI)
     and Zaragoza Center for Advanced Modeling (ZCAM),
     University of Zaragoza,
     E\,-\,50009 Zaragoza, Spain
}

\author{Piero Decleva}
\address{
     Dipartimento di Scienze Chimiche e Farmaceutiche,
     Universit{\`a} di Trieste,
     Via~L.~Giorgieri 1,
     I\,-\,34127 Trieste, Italy
}
\date{\today}

\begin{abstract}
    Ionization and excitation of water molecules in intense laser pulses 
    is studied theoretically by solving the three-dimensional time-dependent 
    electronic Schr\"odinger equation within the single-active-electron 
    approximation. The possibility to image orbital densities by 
    measurement of the orientation-dependent ionization of H$_2$O in 
    few-cycle, 800\,nm linear-polarized laser pulses is investigated. 
    While the highest-occupied molecular orbital $1\,b_1$ is found to 
    dominate the overall ionization behavior, contributions from the 
    energetically lower lying $3\,a_1$ orbital dominate the ionization 
    yield in the nodal plane of the $1\,b_1$ orbital. The ratio of the 
    ionization yields of the two orbitals depends on the intensity. 
    Furthermore, even for laser pulses as long as 8 cycles the 
    orientation-dependent ion yield depends on the carrier-envelope 
    phase. In the interpretation of the orientation-dependent ionization 
    as an imaging tool these effects have to be considered.
\end{abstract}


\maketitle

%
%

\section{Introduction}

Recent efforts to achieve time-resolved imaging 
of chemical reactions or other structural changes with sub-femtosecond 
resolution possess a strong interdisciplinary character, since the goal is  
of interest for physics, chemistry, and biology. Pioneering experiments 
with near-infrared ultrashort laser pulses like the ones in 
\cite{sfm:itat04,sfm:meck08} in which structural information 
about the chemically most relevant valence electrons was extracted from 
the emitted high-harmonic radiation or electron spectra have been of 
great interest, since these approaches should intrinsically have the 
potential for providing also the required time resolution. However, in 
contrast to the initial assumptions, recent experiments have demonstrated 
that the molecular strong-field response depends, at least for some molecules, 
on more than one orbital \cite{sfm:mcfa08,sfm:akag09,sfm:smir09,sfm:farr11a}. 
While such multi-orbital effects clearly complicate simple imaging schemes, 
they can also be the source for even richer information that can be gained 
from such experiments. An example is the electron-hole dynamics in the 
laser-generated ion that may be observed by analyzing the high-harmonic 
radiation \cite{sfm:smir09}. Though very exciting by itself, this  
appears to make direct imaging of the valence electrons and their 
field-free dynamics during, e.\,g., a chemical reaction more complicated. 

The proposed imaging schemes based on linear-polarized ultrashort 
near-infrared laser pulses may roughly be divided into two categories, 
rescattering-based schemes and direct imaging. The first category is 
based on the celebrated three-step model of strong-field physics in which 
(1) an electronic wavepacket leaves the molecule around the local maxima 
of the electric field by tunneling ionization, (2) this wavepacket  
is accelerated in the laser field and reverses its direction as the 
field direction changes, and (3) the electronic wavepacket may 
recollide with its parent ion. As a consequence of this recollision, 
the electronic wavepacket may partly scatter elastically (diffraction) 
or inelastically (leading to excitation or further fragmentation), 
or recombine by the emission of high-harmonic radiation. Clearly, 
all these processes should depend on the structure of the molecular 
ion and thus have the potential for revealing structural information. 
This includes both electronic structure as well as nuclear geometry. 
Corresponding reviews may be found in \cite{sfm:lein07,sfm:lin10}.

\section{Direct imaging using orientation-dependent ionization}
While the rescattering-based imaging schemes are evidently based on 
the third step of the three-step model and consequently require that 
this model is at least a good approximation, the direct imaging 
schemes are based on the first, the ionization step. Measuring the 
ion yield as a function of the time-delay between two very short 
pulses has been demonstrated theoretically \cite{sfm:goll06} and 
experimentally \cite{sfm:ergl06} to induce and image vibrational 
motion in the electronic ground state of neutral molecules. Simplified 
models like the strong-field approximation of the time-dependent 
Schr\"odinger equation suggest also the possibility 
to image the nuclear geometry of a molecule by comparing the 
energy-resolved electron spectra for different molecular orientations, 
but were immediately shown to fail using a full-dimensional 
ab initio treatment for H$_2$ \cite{sfm:vann10}. An alternative, rather 
straightforward approach to time-resolved direct imaging with intense 
near-infrared laser pulses is based on recording the 
dependence of the total ionization probability as a function of 
the relative orientation of the polarization axis of a linear polarized 
laser pulse with respect to the molecule. Since this corresponds to the 
fully integrated electron (or ion) spectrum, it is  
by far the most intense response signal, many orders of magnitude larger 
than recollision signals. This makes it experimentally 
very attractive. In \cite{sfm:pavi07} it was demonstrated that the 
alignment-dependent ionization probability reflects the shape of the 
highest-occupied molecular orbital (HOMO) for both N$_2$ and O$_2$. 
However, the therein also considered example of CO$_2$ turned out to be 
not as straightforward. A possible reason could be a coherent core trapping 
of the valence electron \cite{sfm:petr10a} which explains that only at 
lower intensity the imaging appears to work for CO$_2$, in agreement 
with an experiment performed with lower laser intensity \cite{sfm:thom08}.  

Finally, it should be reminded that time-resolution is achieved in 
direct imaging schemes using pump-probe arrangements. Typically, 
the dynamics has anyhow to be triggered in some way. This may be 
achieved by two identical, time shifted laser pulses as in 
\cite{sfm:goll06,sfm:ergl06} or by synchronizing another pump laser  
to the probe pulse. On the other hand, especially the high-harmonic 
based recollision imaging schemes possess an intrinsic time resolution 
that is, however, limited in total duration to about half a laser 
cycle and thus to about 1.6\,fs for 800\,nm laser light 
\cite{sfm:bake06}. To follow dynamics on longer time scales, 
lasers with longer wavelengths have to be used, or a pump-probe 
scheme has to be applied as in the direct imaging schemes. Furthermore, 
the intrinsic time resolution can only be used for processes that are 
induced in the ionizing first step and thus not really adequate for 
the general purpose of imaging chemical reactions. 
 
In order to image chemical reactions, it is, of course, important 
that concepts tested so far for linear molecules are also 
applicable to more general molecules. We have thus decided to 
investigate the direct imaging proposal based on the recording 
of the orientation-dependent ionization yield for a simple 
but very important non-linear molecule: water. 
Despite its natural importance due to its abundance in nature, 
there are some additional arguments for choosing this molecule 
in the present context. The three energetically highest-lying orbitals 
of H$_2$O are known to very roughly correspond to the three p orbitals 
of the oxygen atom that are, however, differently modified by the 
OH bonds. All three orbitals are thus structurally 
relatively simple and a good test candidate for imaging schemes. 
For example, the energetically highest-lying occupied orbital, 
the $1\,b_1$ HOMO, is almost identical to an oxygen p orbital. 
In contrast to true atomic p orbitals, the two hydrogen atoms in 
water lead, however, to a break of the isotropic symmetry and a 
deformation, especially of the so called $3\,a_1$ \HOmI{}. 
Furthermore, due to the symmetry reduction the three oxygen p orbitals 
loose their degeneracy and thus become (also energetically) 
distinguishable. As a consequence, it is possible to test the 
sensitivity of an imaging scheme. Note, it has to be  
mentioned that the field-free alignment of a dipolar molecule 
like H$_2$O is experimentally a challenging task. However, 
orientation-resolved information for one-photon ionization was 
obtained in \cite{sfm:yama09}, but from a multiple coincidence 
measurement that allowed to reconstruct the orientation at the 
incident of ionization. Other approaches that avoid pre-alignment 
of the molecule are based on the use of circular polarized laser 
pulses as was applied to H$_2$ in \cite{sfm:stau09}. Clearly, 
in this case the theoretical description should consider circular 
and not linear polarization. Due to the increased numerical 
demands, this is, however, so far seldom the case 
(see, e.\,g., the theoretical analysis of the just mentioned 
H$_2$ experiment in \cite{sfm:stau09,sfm:vann10,sfm:jin11a}).

\section{Previous studies of H$_2$O in intense laser fields}
Ionization and harmonic generation of water molecules in intense 
fields have recently stirred some interest and were investigated both 
theoretically and experimentally. The case of relatively low intensities 
with supposedly negligible ionization was studied using time-dependent 
configuration interaction (restricted to single and double excitations) 
in \cite{sfm:krau07} where polarizabilities and low harmonics were 
reported. Ionization of the HOMO of H$_2$O in electric half-cycle pulses 
was studied in \cite{sfm:borb10} within a hydrogenic-orbital approximation 
and thus an atomic model. Adopting a numerical implementation of 
time-dependent density functional theory (TD-DFT) based on Voronoi cells 
the ionization of water molecules in intense laser fields was studied 
in \cite{sfm:son09b}. In that implementation only next-neighbor-cell 
interactions were considered and thus approximate TD-DFT results were 
obtained. This approximate TD-DFT approach is in the following named
TDVFD as in \cite{sfm:son09b}.
Considering two 2-dimensional cuts it was found that within the
TDVFD approach the overall ionization 
is dominated by ionization from the HOMO, but at specific directions 
the \HOmI{} contribution can locally exceed the one from the HOMO. 
In \cite{sfm:zhao11} the ionization from the HOMO of water was 
calculated within the molecular Ammosov-Delone-Krainov (MO-ADK) 
approximation. At least along the considered two-dimensional cut 
the dependence of the ionization rate on the relative orientation 
between laser field and molecular plane as obtained by the MO-ADK 
model was shown to agree qualitatively, though not quantitatively, 
with the TDVFD results in \cite{sfm:son09b}.   

The high-harmonic radiation of water molecules was recently investigated 
experimentally \cite{sfm:wong10} and theoretically within a variant 
of the strong-field approximation in which the recombination dipole 
matrix element is related to single-photon ionization cross-sections 
\cite{sfm:zhao11b}. Already earlier, a theoretical 
investigation within the strong-field approximation was reported 
in \cite{sfm:falg10}. In accordance with the in \cite{sfm:son09b} 
predicted overall dominance of the HOMO ionization (first step of 
the three-step model), the theoretical investigation of isotope effects 
on the high-harmonic 
generation of water was restricted to the HOMO in \cite{sfm:falg10}. It 
was concluded that practically no isotope effect is present and thus 
no nuclear motion is visible from the high-harmonic radiation, if the 
ratio of the high-harmonic spectra from H$_2$O and D$_2$O is analyzed. 
The reason is the absence of (substantial) excitation of vibrational 
motion, if ionization from the water HOMO is considered. This is due 
to the fact that the potential surfaces of the neutral water molecule 
and the formed ion (in its electronic ground state) are very similar. 
The time-resolved imaging of nuclear motion using high-harmonics requires 
on the other hand vibrational excitation and thus different potential 
surfaces of the neutral and the ion 
\cite{sfm:saen00c,sfm:urba04,sfm:goll06,sfm:patc09}.  
Nevertheless, a recent mixed experimental and theoretical study 
\cite{sfm:farr11a} has shown a pronounced isotope effect in  
water harmonics. It is explained by high-harmonic generation from the 
\HOmI{} which, in contrast to the one from the HOMO, is accompanied 
by substantial vibrational excitation. As it becomes clear from the 
present study, the good visibility of the high-harmonic radiation 
from the \HOmI{} (relative to the one from the HOMO) is, within 
the terminology of the three-step model, also due to the second 
and third step, while ionization (in an isotropic ensemble) occurs 
dominantly from the HOMO, at least for not too high intensities.    

The main goal of the present work is, however, to provide quantitative 
predictions of the full three-dimensionally resolved dependence of 
the ionization yield of a water molecule on the relative orientation 
of the molecule with respect to the laser field axis. Based on these 
results the prospects of direct (time-resolved) imaging of orbital 
densities by recording the orientation or alignment dependent 
ionization probabilities are discussed.

%
%

\section{Method}

The full {\it ab initio} treatment of the time-dependent response of 
molecules exposed to intense near-infrared laser fields is a non-trivial 
task. In fact, only very recently it has become possible to 
treat the two electrons of the simplest stable molecule, H$_2$, 
exposed to intense ultrashort laser pulses (with 800\,nm wavelength) 
for arbitrary orientations in full dimensionality \cite{sfm:vann10}. 
Therefore, we have implemented a method that solves 
the time-dependent Schr\"odinger equation for in principle arbitrary 
molecules within the single-active-electron (SAE) approximation. The  
method was introduced in \cite{sfm:awas08} where also the validity 
of the SAE approximation was demonstrated for H$_2$. The extension to 
linear molecules with more than two electrons (N$_2$, O$_2$, and CO$_2$) 
followed in \cite{sfm:petr10a}. As it turned out, the standard 
frozen-core approximation is neither gauge invariant nor does it 
preserve the Pauli principle during the pulse, but it is possible 
to formulate a single-Slater-determinant approximation that does 
\cite{sfm:petr12a}. In the present work, almost all results were 
obtained with the latter approximation (named method B 
in \cite{sfm:petr10a}). Since details of the approach can 
be found elsewhere \cite{sfm:awas08,sfm:petr10a,sfm:farr11a,sfm:petr12a}, 
it is only briefly described with an emphasis on the present 
implementation for general, non-linear molecules.  

\begin{figure}
  \begin{center}
    \includegraphics[width=0.25\textwidth]{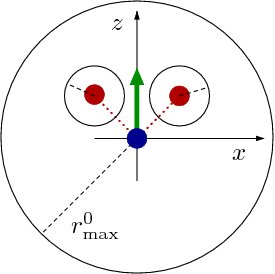}
    \caption{\label{fig:dft_sketch}
    (Color online) Sketch (not to scale) of the spheres defining the 
    local basis sets for H$_2$O. The origin is chosen to coincide with 
    the oxygen atom (blue), while the two H atoms (red) and thus the 
    molecule lie in the $xz$ plane. Two small spheres are centered 
    around the hydrogen atoms, while one large sphere (with radius 
    $r_{\rm max}^0$) defines the main basis and discretization volume.
    The bold green arrow along the positive
    $z$-axis indicates the direction of the permanent dipole moment of water.
    }
  \end{center}
\end{figure}

From the theoretical viewpoint the description of the ionization 
of non-linear molecules is challenging for two reasons. First, 
a proper description of the molecule has to account for its full 
multi-center character. Second, non-linear molecules belong to 
symmetry groups with no infinite symmetry element. 
Therefore, the use of symmetry does not 
lead to a dramatic reduction of the numerical efforts, since the 
Hamiltonian consists of a finite number of blocks and is by far 
not as sparse as for non-Abelian symmetries. The degree of sparseness 
determines on the other hand the efficiency of the time propagation. 
In the case of H$_2$O with $C_{2v}$ symmetry there are only four
irreducible representations: $A_1$, $A_2$, $B_1$, and $B_2$. In view 
of this, compared to linear molecules, small reduction of the 
computational efforts it is evident that the present study demonstrates  
the ability of our recently developed approach 
\cite{sfm:awas08,sfm:petr10a} to treat arbitrary molecules, 
because for the worst case, molecules with no symmetry ($C_1$), 
the number of irreducible representations is only reduced  
from four to one.  

In the first step of the calculation field-free Kohn-Sham orbitals 
are obtained using 
the LB94 exchange-correlation potential which has the correct asymptotic
behavior \cite{gen:vanl94}. The Kohn-Sham orbitals are described by a 
multi-centered $B$-spline basis \cite{bsp:toff02}. 
This basis is composed of a set of (typically atom-centered) spheres 
in which a local basis is defined as the product of spherical harmonics 
and a radial part expressed in $B$ splines. These spheres must 
(presently) not overlap. An 
arbitrary number of such spheres centered at any position of 3D space 
is allowed. Molecular symmetry is 
accounted for by generating symmetry-adapted basis sets.
A large central sphere that is usually 
positioned at the charge center of the molecule is added that defines 
an additional large $B$-spline basis. It overlaps with all 
atomic-centered spheres and provides an improved description of 
the chemical bonding and, especially, the molecular electronic continuum. 
The latter is obtained in a discretized form, the density of the 
discretized continuum being determined 
by the size of the central sphere. The concept is sketched for the 
present example of H$_2$O in Fig.~\ref{fig:dft_sketch}. Since the 
charge center lies very close to the oxygen nucleus, no local basis 
is defined for the latter, but the center of the large sphere 
coincides with the oxygen nucleus and not with the charge center.  
It should be emphasized that the adopted multi-center basis is well 
suited to treat the atomic Coulomb singularities and the atomic 
cores of a molecule. For example, in the frozen-core SAE approach 
in \cite{sfm:abus10} a fitted core potential had to be smoothened 
(by hand) in order to prevent numerical instabilities arising from 
the Coulomb singularities. Furthermore, much higher angular momenta 
are required for describing a multi-center molecule in a single-center 
basis. Also the Cartesian coordinates used in \cite{sfm:span09} 
are not optimal, as the slow quantitative convergence of that 
approach demonstrates.  

\begin{table}
  \begin{center}
    \begin{tabular}{c|cc}
                     &  \multicolumn{2}{c}{$I_p$ (eV)} \\
        Orbital      &Calculated &  Experiment \cite{gen:bann75}\\ \hline
    \rl 1$b_1$       &   13.15   & 12.61  \\
        3$a_1$       &   15.09   & 14.73  \\
        1$b_2$       &   18.69   & 18.55  \\
        2$a_1$       &   30.50   & 32.20  \\
        1$a_1$       &   535.16  & 539.70 \\
    \end{tabular}
    \caption{\label{tab:moldata}
    Calculated vertical ionization potentials ($I_p$)
    for the occupied orbitals of water compared to experimental values.
    }
  \end{center}
\end{table}

In the present calculations, the radius of the large sphere is 
$r_{\mathrm{max}}^0 = 159.5$~a.\,u.\ and the maximum number of angular momenta 
is $l_{\rm max} = 7$. The experimentally determined equilibrium geometry
($R_{\rm OH}$~=~0.958$\mathring{A}$,
 $\varangle$HOH~=~104.5\ensuremath{^\circ} \cite{gen:herz66}) is used and 
the water molecule lies in the $xz$ plane. 
The calculated eigenenergies of the KS orbitals occupied in the ground 
state are given in Tab.~\ref{tab:moldata} and compared to experimental 
values taking the corresponding considerations in \cite{gen:hame02} 
into account.

The time-dependent wavefunction is expressed as a single Slater 
determinant; the initial state being the ground-state Kohn-Sham 
Slater determinant. The one-electron wavefunctions that form 
the time-dependent Slater determinant are expanded in the basis 
of the occupied and virtual Kohn-Sham orbitals of the field-free 
Kohn-Sham molecular Hamiltonian. Thus the Kohn-Sham orbitals form 
a convenient basis, but in principle any other complete single-particle 
basis could be used, like, e.\,g., the Hartree-Fock basis additionally 
used in \cite{sfm:awas08}. It should be emphasized that the present 
approach is not TD-DFT. In contrast, the orbitals are propagated 
in the (frozen) ground-state Hamiltonian plus the external field. 
As is shown in \cite{sfm:petr12a}, 
the time-propagation of the Slater determinant can be reduced to an 
independent time propagation of the initially occupied orbitals. 
The ionization yield is defined as the population of all field-free 
continuum orbitals after the pulse. The total population 
is normalized to 1, an ionization yield of, e.\,g., 0.1 corresponds 
to 10\,\% ionization. As excitation we define the population of all 
field-free, initially unoccupied (virtual), but non-continuum 
orbitals after the pulse. The inclusion of orbitals with an 
energy lower than 20\,a.u.\ led to about 6,000 orbitals that were 
used in the calculations where the laser field is parallel to the $x$, 
$y$, or $z$ axis due to symmetry and dipole-selection rules, while 
about 12,000 orbitals were used for arbitrary orientations. The laser 
pulses were modeled using a cos$^2$ envelope for the vector potential, 
\begin{equation}
     \label{equ:A_cos}
     A(t) = A_0\,\cos^2\left(\frac{\pi t}{T}\right)\sin{(\omega t+\phi)} 
                                                                     \quad ,
\end{equation}
with pulse duration $T$, laser frequency $\omega$, and carrier envelope phase
$\phi$. The ionization yields are multiplied with a factor 2, since all 
orbitals of H$_2$O are doubly occupied (and non-degenerate).  This 
procedure was shown to be appropriate for H$_2$, except for 
very high ionization yields (above about 10 to 20\,\%) where a factor 
1 should be used \cite{sfm:awas08}. 
The calculations were performed in length gauge and within the intervall
$\varphi \in [0, \pi/2]$ and $\vartheta \in [0, \pi]$, where $\varphi$
and $\vartheta$ represent the usual spherical coordinates. The full 
three-dimensional representations are obtained by symmetry arguments.

%
%

\section{Results}

\begin{figure*}[p]
  \begin{center}
    \includegraphics[width=\textwidth]{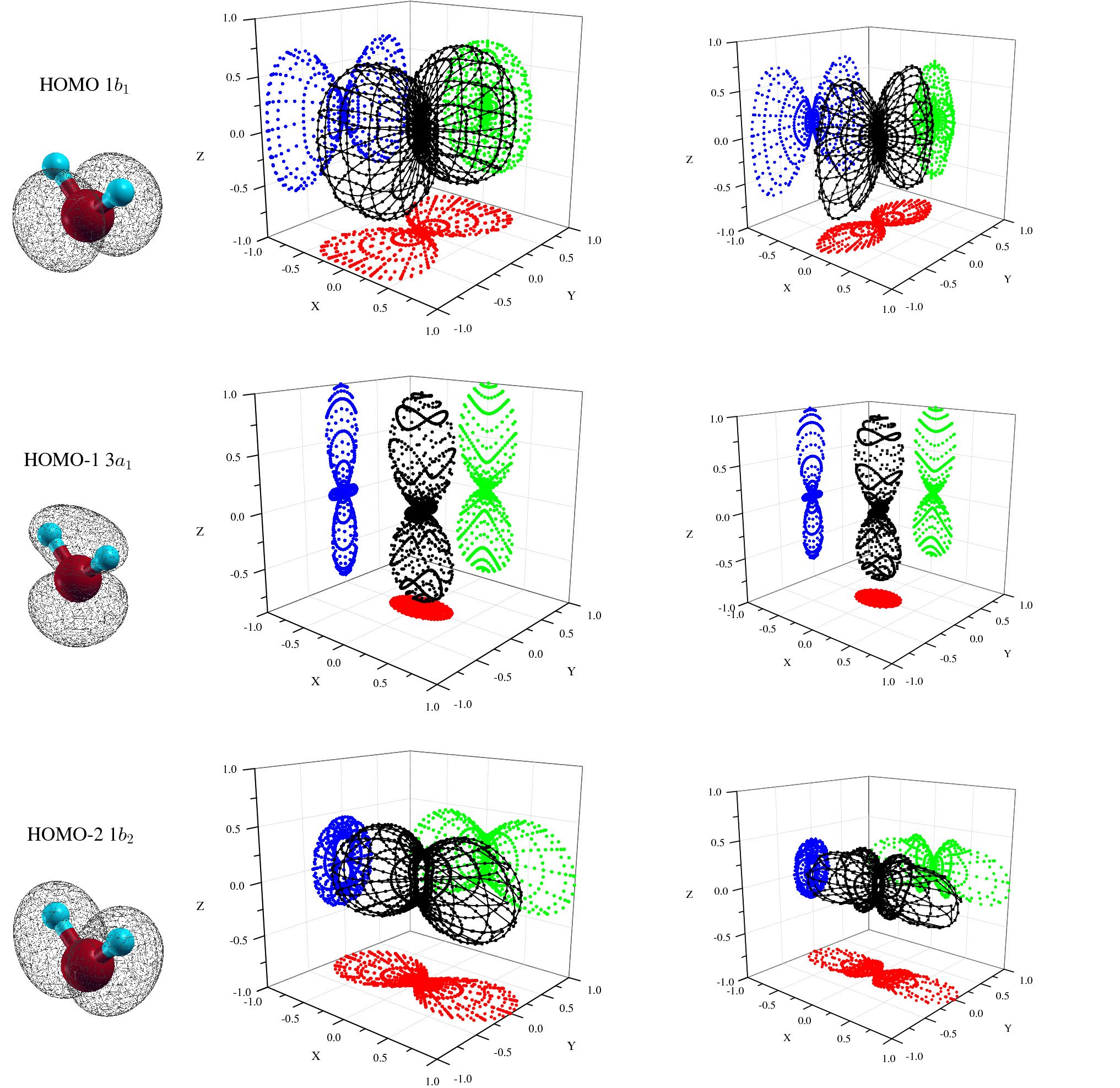}
    \caption{\label{fig:3d_plot}
    (Color online) Three dimensional plots of the normalized 
    ionization (center column) and excitation (right column) yields 
    of the HOMO (top), \HOmI{} (middle) and \HOmII{} (bottom). 
    In each plot the silhouette (maximum circumference) is projected 
    onto the $yz$ (blue), $xz$ (green) and
    $xy$ plane (red) for a better visualization.
    The pulse parameters are  
    a 800~nm, 8~cycle cos$^2$ pulse with peak intensity \intensity[2][13].
    The shape of the individual orbitals (in real space) is given in 
    the left column for comparison.
    }
  \end{center}
\end{figure*}

The shapes (iso-surfaces of identical position-space density) of the three 
energetically highest lying occupied molecular orbitals of H$_2$O are 
visualized in the left panel of Fig.~\ref{fig:3d_plot}. Ordered by 
increasing electron binding energy, the orbitals are denoted as HOMO 
($1\,b_1$), \HOmI{} ($3\,a_1$), and \HOmII{} ($1\,b_2$). The response 
of H$_2$O to an 8-cycle $\cos^2$ pulse (see Eq.~\ref{equ:A_cos}) with a 
central wavelength of 800\,nm, the laser peak intensity \intensity[2][13], 
and $\phi=0$ is also shown in Fig.~\ref{fig:3d_plot}. 
The ionization yields (middle column) and excitation 
yields (right column) are normalized with respect to the 
maxima, i.\,e.\ the maximum yield in each plot is one. In 
agreement with the shapes of the orbitals the ionization 
yields show approximately p-orbital like structures that 
point along the directions of maximal extension of the orbitals. 
This is most evident for the HOMO that also itself is closest 
to an undistorted p orbital of an oxygen atom (the p$_y$ in 
our chosen coordinate system).

\begin{figure}
  \begin{center}
    \includegraphics[width=0.45\textwidth]{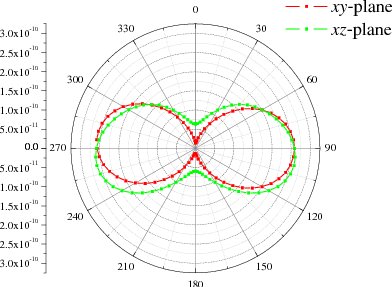}
    \caption{\label{fig:homo-2_xyz}
    (Color online) Comparison of the cuts of the ionization yield
    of the \HOmII{} in $xy$ and $xz$ plane. The color coding agrees
    with the one in Fig.~\ref{fig:3d_plot}. 
    }
  \end{center}
\end{figure}

The \HOmII{} is 
also rather similar to an oxygen p orbital (the p$_x$), but 
the two lobes are distorted in the direction of the hydrogen 
atoms (within the $xz$ plane). The ionization yield of the 
\HOmII{} is not rotationally symmetric around the $y$ axis, 
but shows a similar distortion in the $xz$ plane as the orbital 
itself. This can be seen even more clearly in the corresponding 
cuts shown in Fig.~\ref{fig:homo-2_xyz} in which the absolute 
values of the ionization yield as a function of the orientation 
of the molecule with respect to the field axis is shown. However, 
the distortion is opposite to the one of the orbital, since it 
points toward the negative $z$ axis, while the hydrogen atoms 
point toward positive $z$. In fact, one would not expect any 
asymmetry of this kind in in the strong-field response to a 
linear polarized laser field, since the electric field vector 
points periodically ``up'' and ``down''. Therefore, the total 
ionization probability should be inversion symmetric. The found 
asymmetry is a consequence of the shortness of the adopted laser 
pulse, though 8 cycles is not ultrashort. As a check, a 
calculation with a different carrier envelope phase
$\phi=\pi$ (instead of 0) was performed. As expected, the 
ionization yields are a mirror image of the ones shown in 
Figs.~\ref{fig:3d_plot} and \ref{fig:homo-2_xyz}, i.\,e.\ the 
lobes of the \HOmII{} point for $\phi=\pi$ toward positive 
$z$. The strong phase dependence found here even for a pulse as 
long as 8 cycles is very likely to be a consequence of the strong 
dipole moment of (oriented) water molecules. Thus, a very pronounced 
carrier envelope effect should occur for even shorter laser pulses. 
Since the dipole vector of the molecule is oriented along the 
$z$ direction and the orbital density possesses mirror symmetry 
with respect to the $yz$ and $xz$ planes, the asymmetry occurs 
in the $xz$, but not in the $xy$ plane. While the strong phase 
effect is interesting by itself and provides in sufficiently short 
pulses even information on the absolute orientation of a  
molecule and not only its orientation, it is problematic 
from the viewpoint of orbital imaging. In the case of the 
\HOmII{} of water it could appear as the correct image is obtained, 
but the ``image'' depends on pulse length and carrier envelope 
phase. Recall that in \cite{sfm:akag09} the asymmetry was 
measured for HCl molecules using 50\,fs pulses (and circular 
polarized light), but this required the coincident detection of 
ions in order to restore the molecular-frame information. 
Based on a simplified tunneling model, the found asymmetry 
was attributed to the dipole moment and not to the orbital shape. 
Since a permanent or induced dipole moment influences the 
orientation-dependent ionization yield, this effect has to be 
corrected for, if orbital (densities) should be imaged. The size 
and importance of this correction increases with decreasing pulse 
length and thus, if a better time resolution should be obtained, 
by employing shorter pulses.

\begin{figure}
  \begin{center}
    \includegraphics[width=0.45\textwidth]{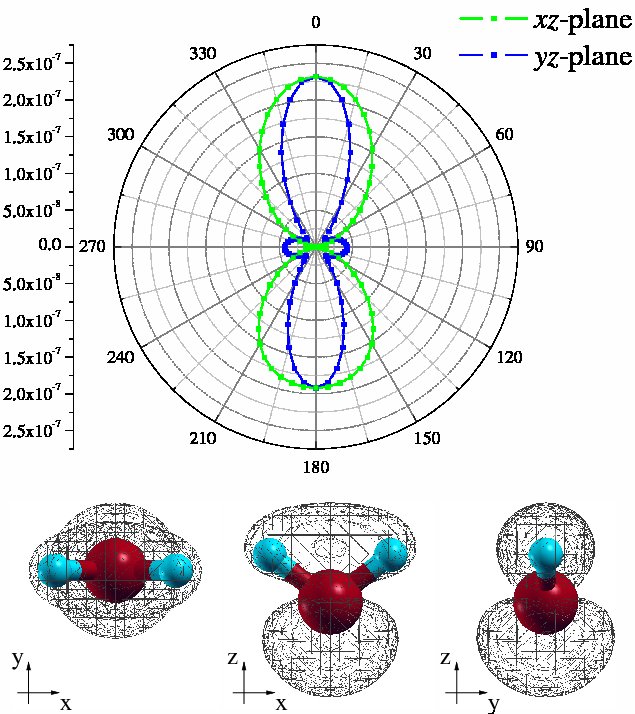}
    \caption{\label{fig:homo-1_xyz}
    (Color online) Shown are cuts of the ionization yield
    of the \HOmI{} in the $xz$ and $yz$ plane. The color coding agrees
    with the one in Fig.~\ref{fig:3d_plot}. 
    Below, the shapes of the \HOmI{} projected on the $xy$, $xz$
    and $yz$ plane are represented.
    }
  \end{center}
\end{figure}

The orientation-dependent ionization of the \HOmI{} shows an 
asymmetry along the $z$ direction similar to the one discussed 
for the \HOmII{}. Correspondingly, the mirror image (with 
respect to the $xy$ plane) is obtained for a phase $\phi=\pi$ 
and the asymmetry disappears for long pulses. Interestingly, 
a second asymmetry is observed when comparing the cuts through 
the $xz$ and $yz$ planes, shown on the absolute scale  
in Fig.~\ref{fig:homo-1_xyz}. The lobes in the $xz$ plane are 
found to be substantially broader than the ones in the $yz$ 
plane. Since the discussed difference in the widths is perpendicular 
to the dipole moment (and shows the proper mirror symmetry), it 
is clearly a consequence of the shape of the orbital. In agreement 
with the orbital shape, the ion yield in the $xz$ plane is broader 
than in the $yz$ plane, since the orbital density of the \HOmI{} 
is stretched in direction of the hydrogen atoms. However, for a 
sufficiently long pulse or an experiment with no control over the 
carrier envelope phase inversion symmetry is automatically enforced 
and thus the widths of the lobes pointing in positive or negative 
$z$ direction will be equal in such a case. The different width  
is thus only a consequence of the upper lobe (pointing toward $+z$) 
of the orbital in Fig.~\ref{fig:homo-1_xyz}, while the lower 
lobe ($-z$) dilutes the effect, since it is rotationally symmetric.
 
This result is of interest, since it confirms directly that the 
orientation-dependent ionization reflects the position 
space orbital and not its momentum space counterpart, since 
position and momentum spaces are reciprocal and a broad distribution 
in position space corresponds to a narrow one in momentum space 
and {\it vice versa}. While this finding is in immediate agreement 
with the position-space based tunnel picture, it seems, on the first 
glance, to contradict the predictions of the strong-field approximation 
(SFA). Especially in its velocity-gauge version it predicts 
(within the frozen-core approximation) that the ionization rate 
of position-space orbital $\phi_j$ is related to its Fourier 
transform $\langle\, {\bf k}_N \,|\, \phi_j \,\rangle$ and 
thus to the momentum-space orbital \cite{sfa:beck05},%
\begin{equation}
        \Gamma_{j}^{\rm (SFA)} \;=\; \sum_{N=N_0}^\infty \:  
                  \int {\rm d}\hat{\bf k}_N \: 
                  f(I,\omega,\epsilon \cdot {\bf k}_N) \:
                  | \,\langle\, {\bf k}_N \,|\, \phi_j \,\rangle \,|^2 
                  \quad .
\end{equation}
The momentum ${\bf k}_N$ of the emitted electron 
depends on the number of absorbed photons $N$ that is limited 
below by the minimum number $N_0$ required to overcome the 
ionization threshold. The function $f$ depends on the 
laser intensity $I$ and frequency $\omega$ as well as on the 
projection of the momentum of the emitted electron on the 
laser polarization vector $\epsilon$. For sufficiently high 
laser intensities, $f$ peaks strongly at parallel emission 
of the electron \cite{sfm:vann10} and the integral over all 
emission directions $\hat{\bf k}_N$ reduces (very) approximately 
to the integrand with $\hat{\bf k}_N\approx\hat{\epsilon}$.  
Furthermore, if the laser frequency is low and the intensity 
sufficiently high, but not (much) higher than the over-the-barrier 
ionization threshold, the function $f$ is strongly peaked at low 
momenta $k_N$. Thus, according to SFA the low-momentum part of the 
electron density is probed. Low momenta correspond, however, to 
the electrons far away from the nuclei, i\,e.\ to the asymptotic 
part of the position-space orbital. Therefore, the predictions 
of the tunneling picture (probing the electron density close to 
the barrier) and the ones from the SFA (probing the low-momentum 
component of the momentum-space orbital) are compatible and 
also in agreement with the present findings for H$_2$O. Finally, 
in all formulations going beyond the frozen-core approximation, 
it is the Dyson orbital which should be more properly employed.  
However, the Dyson orbitals of H$_2$O are predicted 
to resemble the usual orbitals, at least for the 
HOMO, \HOmI{}, and \HOmII{} considered in this work \cite{sfm:seab04}.    
 
Figure \ref{fig:3d_plot} also shows the excitation probability of 
water for the given laser pulse. Again, the overall shape resembles 
the one of atomic p orbitals and is thus similar to the ionization 
yield. This similarity is most pronounced for the \HOmI{}, while 
the HOMO excitation yield does not show the rotational symmetry 
around the $x$ axis, but is compressed in the $y$ direction. The 
\HOmII{} excitation yield shows on the other hand some additional 
structure that is absent in the ionization yield. The basic 
similarity of the orientational dependence of the excitation and 
ionization yields indicate the similarity of the processes leading 
to excitation and ionization as was discussed (for atomic helium) 
in \cite{sfa:nubb08}. A broad distribution of highly excited Rydberg 
states is populated, since (in the semi-classical picture) not all 
electron trajectories reach finally the continuum (``frustrated 
tunnel ionization'').

\begin{figure}
  \begin{center}
    \includegraphics[width=0.45\textwidth]{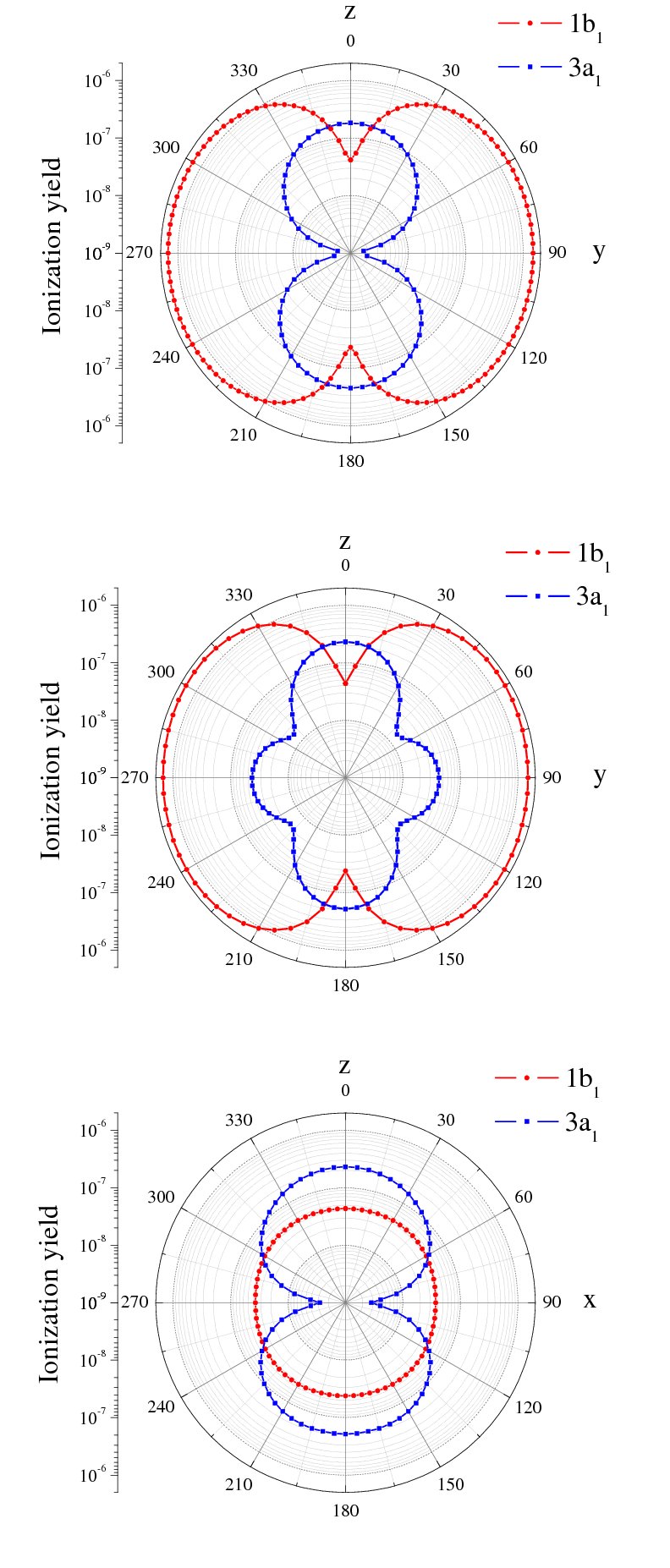}
    \caption{\label{fig:polar_plot_yz}
    (Color online) Comparison of the respective absolute main ionization yields
    of the HOMO $1b_1$ (red) and the \HOmI{} $3a_1$ (blue) in the $yz$ plane
    (top and middle panels) and in the $xz$ plane (bottom panel).
    Laser parameters are the same as in Fig.~\ref{fig:3d_plot}.
    $0^\circ$ mean alignment of the electric field polarization along the $z$ axis
    and $90^\circ$ along the $x$ or $y$ axis.
    The two upper graphs differ by the method applied: Method A (top) and
    method B (middle).
    }
  \end{center}
\end{figure}

The structures in some of the excitation spectra and their absence 
in the ionization yields indicate on the other hand the occurrence 
of resonances that do, however, barely influence the ionization 
process. An exception is the \HOmI{} (small) structure in the $yz$ plane 
that occurs close to the center. It is found in both the excitation 
and the ionization yields. Its origin becomes clearer from the 
comparison of the results that are obtained with the two different 
implementations of the SAE introduced in \cite{sfm:petr10a,sfm:petr12a}. 
The results are shown in Fig.~\ref{fig:polar_plot_yz}. Clearly, 
the structure occurs only in method B in which all orbitals are 
propagated in time, but is absent in method A in which all occupied 
field-free orbitals are excluded in the time propagation. In 
method A the ionization yields resemble in shape very well the one 
of atomic p orbitals. Note, also the small local minimum in the 
ionization yield of the HOMO that occurs along the $y$ axis is 
absent in method A. The reason for these structures seen in 
method B may be interpreted as a coupling between 
the HOMO and the \HOmI. As the HOMO is partially depleted due to 
excitation and ionization, it can partially be refilled from the 
\HOmI. Due to its lower binding energy, this transferred electron 
density is easier excited and ionized which leads to an 
enhanced excitation and ionization yield from the \HOmI. At the same 
time, some population may be transferred from the HOMO into the 
\HOmI{} leading to a stabilization and thus a reduced excitation 
and ionization from the HOMO. A similar coupling of initially 
occupied (field-free) orbitals was found for CO$_2$ \cite{sfm:petr10a}, 
but there the influence on the ionization yield was much more 
pronounced. It should be understood, however, 
that this interpretation is given in terms of field-free 
eigenstates that are simply basis states in the TDSE calculation, 
and thus loose a direct physical meaning during the pulse.  

\begin{figure}
  \begin{center}
    \includegraphics[width=0.45\textwidth]{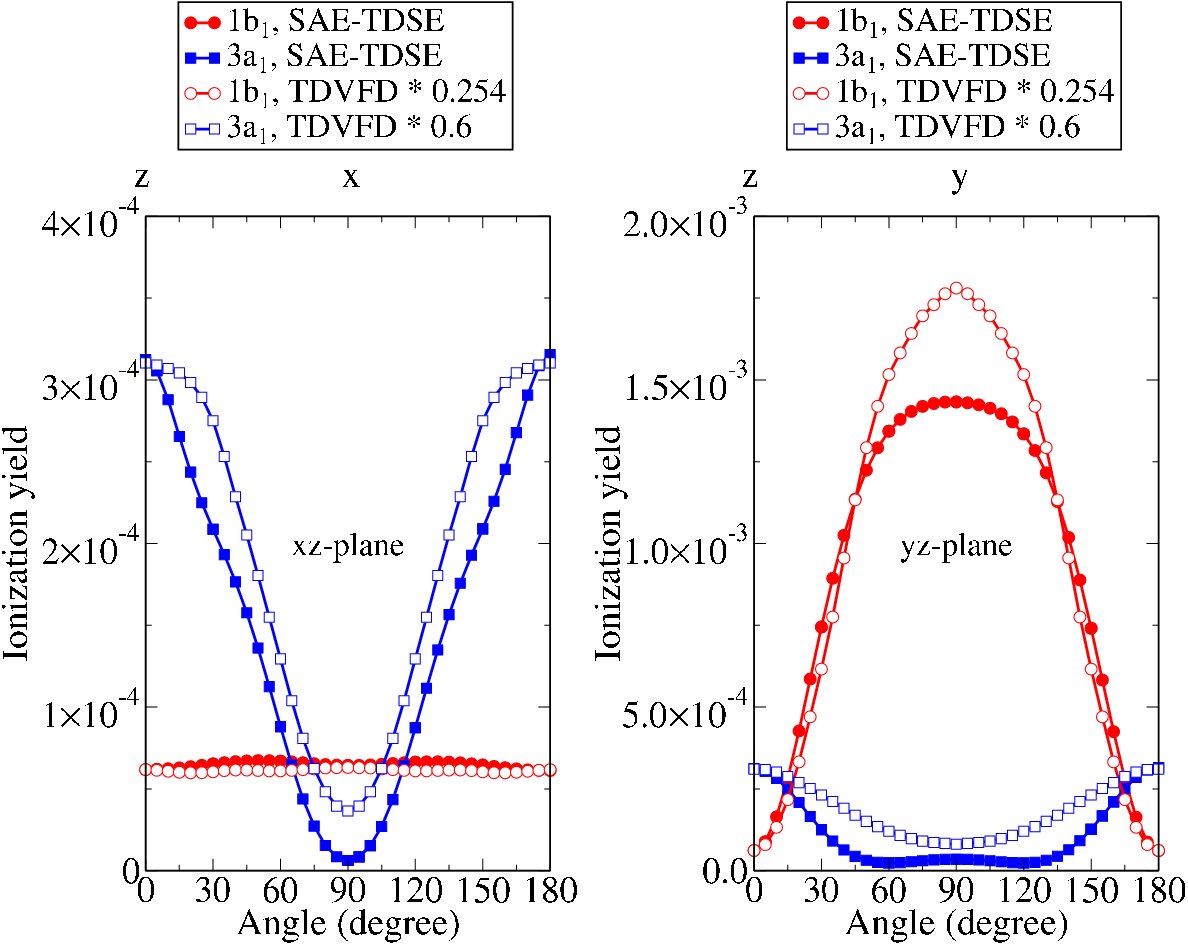}
    \caption{\label{fig:compare_son}
    (Color online) Ionization yields in the $xz$ (left) and $yz$ plane (right)
    for the HOMO and \HOmI{} for a 800\,nm, 20 cycles pulse at the intensity
    \intensity[5][13].  Our SAE-TDSE results are compared to
    TDVFD results \cite{sfm:son09b}. The TDVFD data is scaled (by the factors
    given) to agree with the SAE-TDSE data at 0$^\circ$.
   }
  \end{center}
\end{figure}

For the visibility of the different orbitals in an experiment 
the absolute magnitude of the different ion yields is, of course, 
important. As is evident from Fig.~\ref{fig:polar_plot_yz}, 
within the $xz$ plane the \HOmI{} ionization yield is much 
larger than the one from the HOMO due to the nodal plane of 
the latter. On the other hand, the space volume in which the 
\HOmI{} ionization yield dominates, is rather small, since 
it extends only to about 10\,\% into the $yz$ plane. For the 
laser pulse adopted here, a very well aligned H$_2$O sample 
would be required in order to observe \HOmI{} ionization. 
Integrated over all directions ionization is dominated by the 
one from the HOMO (note the logarithmic scale). A dominant 
ionization from the \HOmI{} in the $xz$ plane had been 
predicted in \cite{sfm:son09b} based on a TDVFD
calculation. A direct comparison (with identical 20 cycle 
pulses with peak intensity \intensity[5][13])  
is shown in Fig.~\ref{fig:compare_son}. Qualitatively, the 
results are very similar, though the present ones tend to 
be broader, i.\,e.\ minima and maxima are slightly less 
pronounced. On the absolute scale, the results differ 
by factors in between 1.7 and 10. 

\begin{figure}
  \begin{center}
    \includegraphics[width=0.45\textwidth]{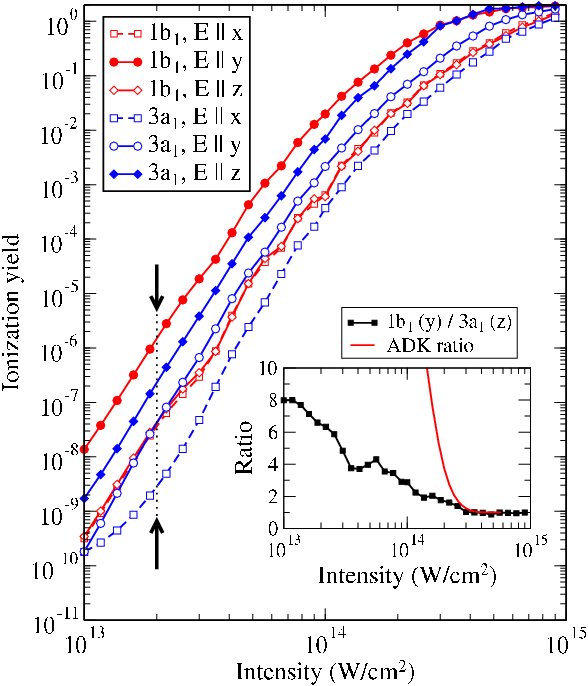}
    \caption{\label{fig:ionization}
    (Color online) Ionization yields of the HOMO $1b_1$ (red) and the \HOmI{}
    $3a_1$ (blue) for alignments of the electric field vector $E$ along the
    three main axes $x$ (squares), $y$ (circles) and $z$ (diamonds) 
    for 8-cycle  cos$^2$ laser pulses (800\,nm). 
    All ionization yields from the SAE-TDSE calculation were multiplied 
    by a factor 2. 
    The inset shows the ratio of the HOMO ionized along the $y$-direction 
    (red filled circles) and the \HOmI{} ionized along the $z$-direction
    (blue filled diamonds) together with an ADK estimate of the ratio.
    The arrows indicate the intensity used in most other graphs, 
    especially in Fig.~\ref{fig:3d_plot}.
    }
  \end{center}
\end{figure}

It is also interesting to consider the intensity dependence 
of the relative ionization yields of the HOMO and the \HOmI{} 
shown in Fig.~\ref{fig:ionization}. For low intensities, the 
maximum ionization occurs along the direction of the HOMO ($y$ axis), 
followed by the one of the \HOmI{} along its maximum extension 
($z$ axis). The ionization of the HOMO along the $x$ and $z$ 
directions is practically identical and, for not too high 
intensities, agrees also to the one of the \HOmI{} along the 
$y$ direction. The lowest probability is found for the 
\HOmI{} along the $x$ direction, due to the absence of the 
resonant structure observed in $y$ direction and discussed above. 
Comparing the ionization probabilities of the HOMO and 
the \HOmI{} along the directions of maximum ionization 
(inset of Fig.~\ref{fig:ionization}) one finds a pronounced 
intensity dependence. Clearly, at high intensities the ratio 
approaches 1, since the HOMO ionization saturates earlier than 
the one of the \HOmI{}. In a comparison to experiment, saturation 
requires to consider also focal volume effects. Furthermore, the 
transition from a factor 2 to 1 due to saturation of the 
single-electron ionization should differ for HOMO and \HOmI{}, if 
it follows the earlier findings for H$_2$ where a rather sharp 
transition between factors 1 and 2 for about 10$\,\%$ ionization 
probability was found. 

In a recent experiment the high-harmonic spectra of H$_2$O and 
D$_2$O have been measured. In the analysis based on a calculation 
using the present theoretical approach it was concluded that the 
harmonics for the different isotopes differ significantly, because 
ionization from the \HOmI{} excites a vibrational wavepacket, while 
HOMO ionization does not \cite{sfm:farr11a}. In view of the present 
finding it is clear that the observability of the \HOmI{} high 
harmonics in an isotropic sample requires that the lower ionization 
probability is partially compensated for by the higher recombination 
rate (3rd step of the three-step model).
Furthermore, as is also 
seen from the present results, saturation plays an important role 
for higher intensities, since the ratio of the maxima of the ionization yields 
from the HOMO and the \HOmI{} approaches 1 for high intensities (cf. the
inset of Fig.~\ref{fig:ionization}). An estimate of this ratio with ADK
(including the saturation effect) is also shown. Clearly, ADK overestimates 
the ratio significantly for lower intensities and reaches the ratio 1 
due to saturation very abruptly.
Concerning the direct imaging scheme discussed above, the intensity 
dependence of the ratio is important, since the image of the HOMO 
density may be distorted by an admixture of lower lying orbitals 
and this admixture changes with intensity. 
While this is very critical for a measurement at a single intensity, 
the intensity dependence of the ratio can, on the other hand, also 
help to disentangle the contributions from different orbitals. 

\section{Conclusion}
In conclusion, the response of water molecules to ultrashort intense 
laser pulses (at 800\,nm) was investigated by solving the time-dependent 
Schr\"odinger equation within a single-active-electron approximation, 
but taking into account a complete multicenter molecular core. In fact, 
the influence of the laser field on all electrons was considered, but 
the electron-electron interaction is restricted to a single-determinant 
approximation. It was found that the orientation-dependent ionization 
from the three energetically highest occupied orbitals reflects their 
overall shape. However, for a sufficiently long pulse the image   
acquires inversion symmetry, since the electric field vector 
points approximately symmetrically up and down. For short pulses, in 
the present case even as long as 8 cycles, deviations from the inversion 
symmetry of the ionization distribution were found. Although it may 
be a consequence of the asymmetric shape of the orbital, the visibility 
of this effect even for such relatively long pulses suggests that it 
is mainly due to the permanent dipole moment of an (oriented) water 
molecule. While this effect provides additional information, it distorts 
the image of the orbitals and it depends evidently on the absolute phase 
and time duration of the laser pulse. On the other hand, the missing 
rotational symmetry of water compared to the previously mainly discussed 
linear molecules allowed a distinction of broad and narrow features of 
the orbitals, provided they survive the enforced inversion symmetry.

The authors would like to thank S.-K. Son for kindly providing their data
in numerical form and COST {\it CM0702} for financial support.
SP, AS, and PD acknowledge financial support within the EU Initial 
Training Network (ITN) CORINF, 
SP and AS from the {\it Fonds der Chemischen Industrie},
AC from the Spanish MICINN (grant FIS2009-13364-C02-01), and 
PD by CNR-INFM Democritos and INSTM Crimson.

\bibliographystyle{model1-num-names}

\end{document}